\documentclass[prl,twocolumn,showpacs,amsmath,amssymb,amsfont,floatfix]{revtex4}

\usepackage{graphicx}

\begin{document}

\title{Measurement of the decay of Fock states in a superconducting quantum circuit}

\author{H. Wang}

\author{M. Hofheinz}

\author{M. Ansmann}

\author{R. C. Bialczak}

\author{E. Lucero}

\author{M. Neeley}

\author{A. D. O'Connell}

\author{D. Sank}

\author{J. Wenner}

\author{A. N. Cleland}
\email[]{anc@physics.ucsb.edu}

\author{John M. Martinis}
\email[]{martinis@physics.ucsb.edu}

\affiliation{Department of Physics, University of California, Santa Barbara, California, CA 93106}

\date{\today}

\begin{abstract}
We demonstrate the controlled generation of Fock states with up to 15 photons in a microwave
coplanar waveguide resonator coupled to a superconducting phase qubit.  The subsequent decay of the
Fock states, due to dissipation, is then monitored by varying the time delay between preparing the
state and performing a number-state analysis. We find that the decay dynamics can be described by a
master equation where the lifetime of the $n$-photon Fock state scales as $1/n$, in agreement with
theory. We have also generated a coherent state in the microwave resonator, and monitored its decay
process. We demonstrate that the coherent state maintains a Poisson distribution as it decays, with
an average photon number that decreases with the same characteristic decay time as the one-photon
Fock state.
\end{abstract}


\maketitle

Fock states are the eigenstates of the harmonic oscillator, and as such play a fundamental
role in quantum communication and information.  Due to the linear nature of the harmonic
oscillator, transitions between adjacent Fock states are degenerate, making it
impossible to generate and detect such states with classical linear techniques. Interposing a
non-linear quantum system, such as that provided by a two-level qubit, allows such states to be
probed unambiguously \cite{meek1996,cira1993,varc2000,bert2002,waks2006,guer2007,
wall2004,hofh2008,joha2006,houc2007,sill2007,maje2007,schu2007,asta2007,fink2008}. However, the
\emph{deterministic} generation and characterization of Fock states with arbitrary photon numbers
has not been demonstrated until recently \cite{hofh2008}, using a superconducting phase qubit
coupled to a microwave resonator. The exquisite control provided by this implementation allows new
tests of the fundamental physics of Fock states.

In this Letter, we present the first experimental measurement of the time decay of Fock
states in a superconducting circuit. Although measurements of the relaxation dynamics have been done for the 2-phonon Fock state
of a trapped ion coupled to a zero-temperature amplitude reservoir~\cite{turc2000}, and proposed
for higher $n$ Fock states of photons by performing a statistical analysis of a sequence of quantum
non-demolition measurements \cite{guer2007}, a direct measurement beyond $n = 2$ has not been
attempted to date. The experiment described here is performed with a system similar to that in Ref.
\cite{hofh2008}, but using a microwave resonator with significantly better performance, allowing
the generation of Fock states with up to 15 photons. The subsequent decay of the Fock states, as
well as the decay of a coherent state generated by a classical microwave pulse, have been directly
measured and shown to display the expected dynamics. More importantly, we
confirm that the lifetime of an $n$-photon Fock state accurately scales as $T_n = T_1/n$, where $T_1$ is the
lifetime for the $n=1$ Fock state \cite{lu1989}. The lifetime $T_n$ scales as $1/n$ because
each of the $n$ photons can independently decay.

In our experiment, the Fock states are generated by first driving the phase qubit from the ground
$\left|g\right\rangle$ to the excited $\left|e\right\rangle$ state, during which the qubit and
resonator are tuned out of resonance. We then turn on the coupling between the qubit and resonator
by frequency-tuning them into resonance, allowing the qubit to transfer the photon to the
resonator, generating an $n=1$ photon Fock state in the resonator and leaving the qubit in its
$|g\rangle$ state. The qubit is then de-tuned from the resonator, so that the subsequent photon
dynamics are determined by the resonator parameters alone. Because this sequence can be repeated
$n$ times to produce an $n$-photon Fock state, this method is scalable to arbitrary photon numbers,
limited only by the resonator decoherence time, the photon transfer speed through the
qubit-resonator coupling, and the fidelities of the qubit $\left|e\right\rangle$ state preparation
and the qubit-resonator tuning pulse. A resonator with a relatively long $T_1$ is thus desirable
for both preparing high $n$ Fock states and measuring the subsequent decay dynamics.

\begin{figure}[t]
\begin{center}
\resizebox{0.48\textwidth}{!}{
\includegraphics[clip=True]{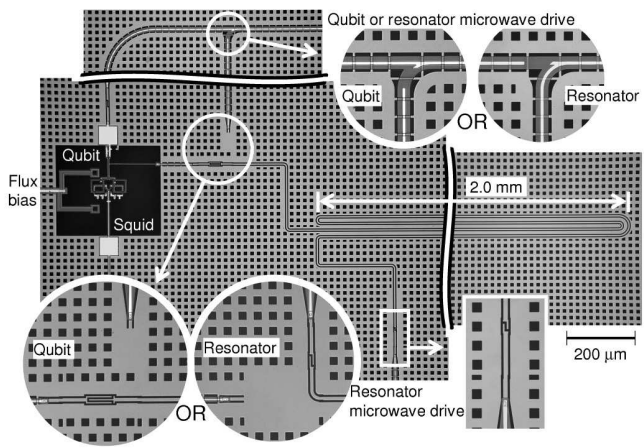}}
\end{center}
\caption{\label{fig.imag} Photomicrograph of  a coplanar waveguide resonator coupled to a phase
qubit. The total length of the half-wavelength resonator is 8.8 mm.  The insets show the two
designs of the microwave lines that were varied in fabrication to connect the resonator either to a
qubit or a classical preamplifier, allowing qubit-based $T_1$ or classical resonator $Q$
measurements, respectively. }
\end{figure}

Our device is composed of a half-wavelength coplanar waveguide resonator, capacitively coupled to a
superconducting phase qubit, as shown in Fig. \ref{fig.imag}. It has a layout similar to that used
in previous experiments \cite{neel20081,hofh2008}.  In order to improve the resonator lifetime
$T_1$, we have implemented a new design and fabrication procedure, that reduces the amount of
amorphous dielectric in the vicinity of the resonator, thereby reducing microwave dielectric loss
\cite{ocon2008}. We also used a resonator fabricated from superconducting Re (with $T_c \approx
1.8$ K), as this metal has substantially less native oxide than Al, which was used previously. As
depicted in Fig.~\ref{fig.imag}, the fabrication incorporates two resonator designs on the same
wafer, allowing us to measure the resonator quality factor $Q$ from a classical resonance
measurement \cite{ocon2008}, as well as the resonator $T_1$ using a qubit-based measurement
\cite{hofh2008}. The layout of the two resonator designs are as identical as possible, to match any
dissipation due to (unwanted) radiating modes in the microwave circuit.

\begin{figure}[t]
\begin{center}
\resizebox{0.48\textwidth}{!}{
\includegraphics[clip=True]{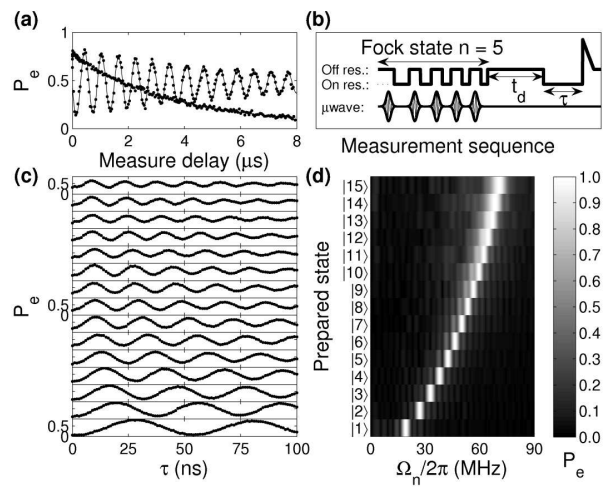}}
\end{center}
\caption{\label{fig.perf} (a) Measurements of energy  decay and Ramsey fringes for the resonator.
Lines are fits to the data, yielding a resonator $T_1'= 3.47\ \mu\textrm{s}$ and $T_2' = 6.53\
\mu\textrm{s}$. (b) Pulse sequence used for the experiment, showing both the flux bias (top) and
microwaves (bottom) needed to generate the $n=5$ Fock state.  (c) Experimental data showing the
state probability $P_e$ for the excited state $\left|e\right\rangle$ versus the qubit-resonator
interaction time $\tau$ during a Fock state readout (see text). The decrease in the qubit
oscillation period from bottom to top, scaling as $\sqrt{n}$, is expected from the
Jaynes-Cummings model for an $n$-photon Fock state. For each panel, the y-axis ranges from
0\% to 85\%. The purity of the sinusoidal oscillations
indicates the high fidelity of the state preparation. (d) Fourier transform of the data in (c). For
each Fock state, the Fourier amplitude (gray scale) is normalized to the maximum peak amplitude.}
\end{figure}

The qubit is characterized using pulse sequences described previously \cite{luce2008}.  We found an
energy relaxation time $\approx 600\, \textrm{ns}$, a phase coherence time $\approx 100\,\textrm{ns}$, 
and a measurement visibility between the qubit $\left|g\right\rangle$ and
$\left|e\right\rangle$ states of close to 90\%.  The initial characterization of the microwave
resonator was made using a pulse sequence that swaps a photon into and out of the resonator, as
described above and in Ref.~\cite{neel2008}.  The decay of the resonator state  measured in this
manner is plotted in Fig.~\ref{fig.perf}(a), showing a decay time $T_1'= 3.47\ \mu\textrm{s}$ that
is three times longer than in our previous experiment \cite{hofh2008}. The decay time for a Ramsey
sequence gives $T_2'= 6.53\ \mu\textrm{s}$, approximately twice the resonator $T_1'$, which
indicates a long intrinsic dephasing time for the resonator, $T_\phi' \gg T_1'$, as expected.  In
addition, the classical measurement \cite{ocon2008} of the resonator quality factor $Q \approx
130,000$ at low power predicts $T_1=Q/\omega_r = 3.15\ \mu\textrm{s}$, consistent with the qubit $T_1$
measurement. Here, $\omega_r /2\pi = 6.570\ \textrm{GHz}$ is the resonator oscillation frequency,
in agreement with design calculations that include the kinetic inductance of the Re center
conductor. A spectroscopy measurement \cite{hofh2008} yields a splitting size of
$\Omega/2\pi = 19.0\pm 0.5\ \textrm{MHz}$, as expected for
the qubit-resonator coupling strength due to a design value of
coupling capacitor of $2.6\ \textrm{fF}$. The qubit-resonator coupling can be turned off by tuning
the qubit off-resonance. In this experiment the detuning was $400\ \textrm{MHz}$.

The pulse sequence for generating and measuring Fock states \cite{hofh2008} is illustrated in
Fig.~\ref{fig.perf}(b).  The timing of the pulses is based on the Jaynes-Cummings
model~\cite{jayn1963}, which predicts that on resonance, the occupation of the two coupled
degenerate states $\left|g,n\right\rangle$ and $\left|e,n-1\right\rangle$, with $n$ and $n-1$
photons in the resonator, respectively, will oscillate with a radial frequency $\Omega_n = \sqrt{n}
\Omega$. For generation, the $n^{\rm{th}}$ photon is loaded from the excited state of the qubit
using a carefully controlled interaction time of $\pi/\Omega_n = \pi/\sqrt{n} \Omega$. The
subsequent readout of the resonator state is performed by tuning the qubit, initially in its ground
state, into resonance with the resonator, and measuring the subsequent evolution of the qubit
excited state probability $P_e$ as a function of the interaction time $\tau$. A Fourier transform
(FT) of $P_e(\tau)$ gives the occupation probability of the $n$-photon Fock state as the Fourier
component (in amplitude) of $P_e$ at the frequency $\Omega_n$.

Figure~\ref{fig.perf}(c) shows $P_e$ versus interaction time for a number of different Fock states.
For these data, the readout starts after a delay time $t_d = 150\,\textrm{ns}$.  For an interaction
time $\tau$ from 0 to 300 ns, sinusoidal oscillations are clearly visible for Fock states from
$n=1$ up to  $n$ = 15. The visibility of the $n=1$ sinusoidal oscillation is about 75\%, which results from 
the high fidelity of the state preparation and the high visibility of qubit readout close to 90\%. 
The frequency increases with increasing $n$, consistent with the predicted
$\sqrt{n}$ dependence. A FT of the data is shown in Fig.~\ref{fig.perf}(d), indicating the spectral
purity of the oscillations. The Fock state $n = 1$ gives a Rabi frequency of 19.33 MHz, consistent
with the splitting obtained from spectroscopy.

\begin{figure}[t]
\begin{center}
\resizebox{0.48\textwidth}{!}{
\includegraphics[clip=True]{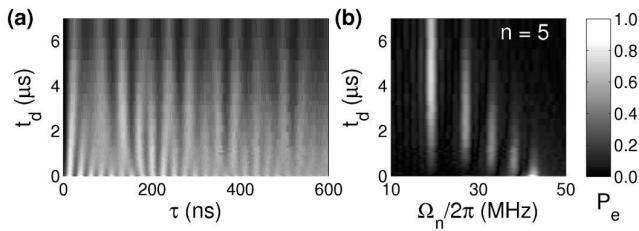}}
\end{center}
\caption{\label{fig.fd2d} (a) Qubit $\left|e\right\rangle$ state probability $P_e$ (gray scale) as a function of
the qubit-resonator interaction time $\tau$ (horizontal axis) and the measure delay $t_d$ (vertical
axis). (b) Fourier transform (along horizontal axis) of data in (a), showing decay of the Fock
states with increasing time.  The Fourier spectrum is normalized to the maximum peak amplitude
corresponding to $n = 5$ at $t_d = 0$.}
\end{figure}

The decay dynamics of the Fock states can be probed by varying the measurement delay $t_d$ in the
pulse sequence.  An example is shown in Fig.~\ref{fig.fd2d}(a), where $P_e$ is plotted in gray
scale versus $t_d$ and $\tau$ for the Fock state $n=5$. At $t_d = 0$, a periodic oscillation of
$P_e$ is observed that corresponds to the occupation of the $n=5$ Fock state. With increasing
$t_d$, the oscillation evolves into a complex aperiodic pattern, finally changing to a slower
oscillation frequency for $t_d \approx 7~\mu\textrm{s}$.  The corresponding FT analysis is shown in
Fig.~\ref{fig.fd2d}(b). The system starts in the Fock state $n = 5$, and then evolves to a mixed
state with several frequencies present in the spectrum, and eventually decays to a mixture of the
Fock states $n = 1$ and $n = 0$.

\begin{figure}[t]
\begin{center}
\resizebox{0.48\textwidth}{!}{
\includegraphics[clip=True]{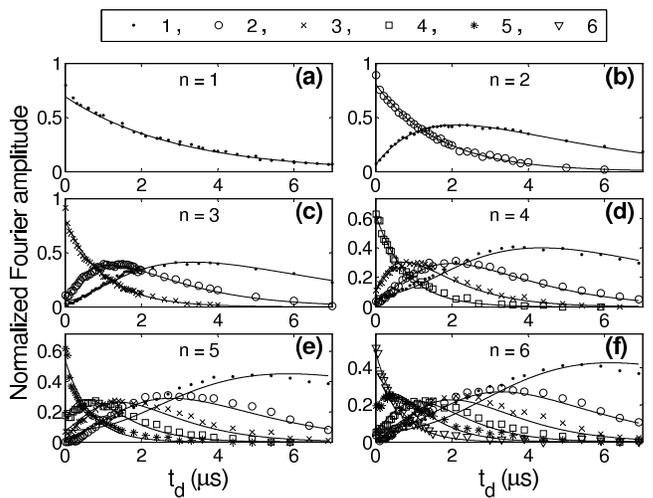}}
\end{center}
\caption{\label{fig.fd1d} Time decay of Fock states from $n$ = 1 (a) to $n$ = 6 (f), taken from the amplitude of the
FT data. Occupation probabilities of the Fock states, $P_n$, are plotted with symbols defined at top panel.
Data are normalized by choosing the total occupation probability, including $n = 0$, at the measure delay $t_d = 0$ to be unity~\cite{hofh2008}. Lines are fits to the data; the $n$-state lifetimes $T_n$ as extracted from the fits are
listed in Table~\ref{tab.T}.}
\end{figure}

As no phase information is involved, the decay of Fock states can be simply described by changes in
the state probabilities.  The master equation is given by
\begin{equation}
\label{eq.mast}
\frac{dP_n(t)}{dt} = -\frac{P_n(t)}{T_n}+\frac{P_{n+1}(t)}{T_{n+1}},
\end{equation}
where $P_n(t)$ and $T_n$ are the occupation probability and the decay time of the $n$-photon Fock
state, respectively. To compare our data with theory, we plot in Fig.~\ref{fig.fd1d}(e) the
normalized amplitude in the FT as a function of the measure delay $t_d$ for each Fock state in the
spectrum of Fig.~\ref{fig.fd2d}(b).  Also shown in Fig.~\ref{fig.fd1d} is the decay of Fock states
from $n$ = 1 (a) to $n$ = 6 (f), similarly obtained from the FT spectrum.  The
predictions from the master equation (lines) fit the data remarkably well for Fock states up to
$n=6$, taking as initial conditions the measured state probabilities.  The fitting parameters for
$T_n$ for each panel are listed in Table~\ref{tab.T}. The average values show good scaling with the
predicted $T_n \approx T_1/n$ for $n = 1$ to 5.  Since 600 ns-long time traces are used for the FT
analysis to retrieve the Fock state probability, it is reasonable to begin to see some deviation
from the prediction for $T_6 \approx 640\ \textrm{ns}$.

\begin{table}[b]
\caption{\label{tab.T} Lifetimes (in units of $\mu\textrm{s}$) of the $n$-photon Fock states taken
from fits to Fig.~\ref{fig.fd1d}(a)-(f), respectively.  The averages are listed in the
second-to-last row, along with the expected $T_1/n$ scaled from the measurement of $T_1 = 3.20\
\mu\textrm{s}$. The values in parentheses are not included in the averages, as the fits are only
reliable if there is a sizeable exponential decay of the data at times less than 7 $\mu\textrm{s}$,
a limit set by our electronics. }
\begin{ruledtabular}
\begin{tabular}{ccccccc}
n     & $T_1$ & $T_2$ & $T_3$ & $T_4$ & $T_5$ & $T_6$ \\\hline
1     & 2.96  &    &    &    &    &    \\
2     & 3.21  & 1.65  &    &    &    &    \\
3     & 3.44  & 1.58  & 1.01  &    &    &    \\
4     & (4.15)  & 1.60  & 1.02  & 0.81  &    &    \\
5     & (6.29)  & (1.88)  & 1.11  & 0.79  & 0.65  &    \\
6     & (5.49)  & (1.78)  & 1.08  & 0.77  & 0.66  & 0.64  \\\hline
Avg.   & 3.20 & 1.61 & 1.06 & 0.79 & 0.66 & 0.64 \\
$T_1/n$ &3.20 &1.60 &1.07 &0.80 &0.64 &0.53 \\
\end{tabular}
\end{ruledtabular}
\end{table}

\begin{figure}[t]
\begin{center}
\resizebox{0.45\textwidth}{!}{
\includegraphics[clip=True]{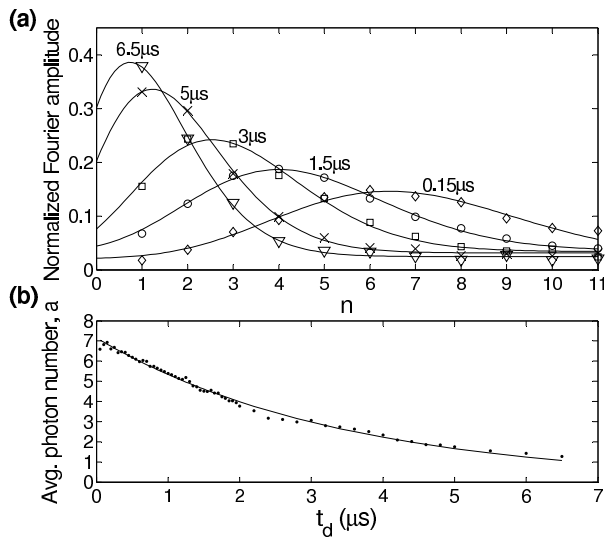}}
\end{center}
\caption{\label{fig.cdec} (a) Occupation probabilities for coherent state analysis, taken from the normalized Fourier amplitude 
versus Fock state number $n$ at a few example
delay times. Data are normalized by choosing the total occupation probability at the measure delay $t_d = 0$ to be
unity, assuming that the $n=0$ probability at $t_d=0$ is negligible. 
Lines are fits to the data according to the Poisson distribution.  (b)
Average photon number $a$ versus the measure delay $t_d$, obtained from fits to data as
shown in (a). The line is an exponential fit yielding a
lifetime of $3.43\ \mu\textrm{s}$. }
\end{figure}

The FT analysis can also be used to test the decay dynamics of a coherent state.  A coherent state
is a superposition of Fock states, with the occupation probability $P_n$ of the Fock state $n$
given by a Poisson distribution,
\begin{equation}
\label{eq.pois}
P_n(a) = \frac{a^ne^{-a}}{n!},
\end{equation}
where $a$ is the average photon number. The resonator is first driven with a 100 ns classical
microwave pulse, creating a coherent state with $a$ proportional to the drive pulse amplitude
\cite{hofh2008}. After a measure delay $t_d$, the resonator state is read out using the qubit,
initially in its ground state, in the same manner as discussed above. For a coherent state with
$\langle n \rangle = a \approx 7.14$, the occupation probabilities of the Fock states versus the
photon number $n$ are plotted in Fig.~\ref{fig.cdec}(a) for several time delays. At each delay time $t_d$, the Fock state
probabilities (points) can be well described by the Poisson distribution given by Eq.~\ref{eq.pois}
(lines), with $a$ as the fitting parameter.  In Fig.~\ref{fig.cdec}(b), we plot $a$ as a function of
$t_d$ obtained for the complete data set. 
The average photon number $a$ decays with a dependence that is
fit by an exponential with a decay time $3.43~\mu\textrm{s}$, in good agreement with the measured
resonator $T_1$.

We note that although the occupation probabilities of the Fock states still follow a Poisson
distribution during the entire decay process, our analysis cannot distinguish a statistical mixture
from a pure coherent state. A complete tomographic measurement is necessary in order to verify the
phase coherence of the coherent state.

In conclusion, we have improved the decay time of our earlier resonator design by a factor of 3,
and thereby generated high-fidelity Fock states with up to 15 photons. The time decay of both
Fock and coherent states were directly measured, and shown to be in excellent agreement with the
theoretical prediction of a master equation with $T_n = T_1/n$.

\textbf{Acknowledgements.} Devices were made at the UCSB Nanofabrication Facility, a part of the
NSF-funded National Nanotechnology Infrastructure Network. This work was supported by IARPA under
grant W911NF-04-1-0204 and by the NSF under grant CCF-0507227.

\end{document}